\documentclass[12pt]{article}
\usepackage{epsfig}
\def\gsim{\;\raise0.3ex\hbox{$>$\kern-0.75em\raise-1.1ex\hbox{$\sim$}}\;}
\def\lsim{\;\raise0.3ex\hbox{$<$\kern-0.75em\raise-1.1ex\hbox{$\sim$}}\;}
\newcommand{\be}{\begin{equation}}
\newcommand{\ee}{\end{equation}}
\newcommand{\bea}{\begin{eqnarray}}
\newcommand{\eea}{\end{eqnarray}}
\newcommand{\bt}{\begin{tabular}}
\newcommand{\et}{\end{tabular}}
\newcommand{\ba}{\begin{array}}
\newcommand{\ea}{\end{array}}

\begin{document}
\setlength{\unitlength}{1mm} {\hfill
    $\ba{r}
    \mbox{Auger-GAP-2003-013} \\
    \mbox{DSF 06/2003} \\
    \mbox{IFIC/03-07}
    \ea$}\vspace*{1cm}

\begin{center}
{\Large \bf Corsika+Herwig Monte Carlo Simulation of Neutrino
Induced Atmospheric Air Showers}
\end{center}

\bigskip\bigskip

\begin{center}
{\bf M. Ambrosio$^1$}, {\bf C. Aramo$^1$}, {\bf A. Della
Selva$^1$}, {\bf G. Miele$^1$}, {\bf S. Pastor$^2$}, {\bf O.
Pisanti$^1$}, and {\bf L. Rosa$^1$}
\end{center}

\vspace{2cm}

\noindent
{\it
\bt{rl}
$^1$ & Dipartimento di Scienze Fisiche, Universit\`{a} di Napoli ``Federico
II'' and \\
& Istituto Nazionale di Fisica Nucleare, Sezione di Napoli \\
& Complesso Universitario di Monte S. Angelo,  Via Cinthia, I-80126 Napoli,
Italy \\
$^2$ & Instituto de F\'{\i}sica Corpuscular (CSIC-Universitat de
Val\`{e}ncia),\\ & Edificio Institutos de Investigaci\'{o}n, Apdo.\ 22085,
E-46071 Valencia, Spain\et \\
}

\bigskip\bigskip\bigskip

\noindent
E-mails: \\
{\bf ambrosio@na.infn.it \\
aramo@na.infn.it \\
dellaselva@na.infn.it \\
miele@na.infn.it \\
pastor@ific.uv.es \\
pisanti@na.infn.it \\
rosa@na.infn.it}

\bigskip\bigskip\bigskip

\begin{abstract}
High-energy neutrino astronomy represents an open window both on
astrophysical mechanisms of particle acceleration and on
fundamental interactions. The possibility of detecting them in
large earth-based apparatus, like AUGER, AMANDA, ANTARES, is quite
challenging. In view of this, the capability of generating
reliable simulations of air showers induced by neutrinos is
mandatory in the analysis of experimental data. In this paper we
describe preliminary results towards the development of a new
version of the Monte Carlo CORSIKA, capable of handling neutrinos
too as primary particles.  In our approach the first interaction
of the primary neutrino is simulated in CORSIKA with a call to the
HERWIG event generator.
\end{abstract}

\thispagestyle{empty} \setcounter{page}{0}

\newpage
\baselineskip=.8cm


\section{Introduction}

A large variety of astrophysical and exotic sources are expected to
emit ultra-high energy (UHE) particles and among them also neutrinos.
The observation of Extensive Air Showers (EAS) \cite{exp} with energy
larger than $10^{20}$ eV implies in any theoretical scenario the
simultaneous presence of a flux of UHE $\nu$'s. This is particularly
clear reminding that such energetic events are either originated in an
astrophysical source (bottom-up models) or are the consequence of the
decay of massive Big Bang relic particles (top-down scenarios)
\cite{sigl}. The bottom-up acceleration mechanisms occur in
astrophysical environments, such as Active Galactic Nuclei or Gamma
Ray Burst sources, characterized by a large product of the magnetic
field times the acceleration length. They are able to produce jets of
hadronic matter at least up to $10^{21}$ eV, and this huge
acceleration of nucleons yields a photo--production of pions and thus
a copious emission of neutrinos carrying some percent of the primary
particle energy. According to the peculiar properties of the
astrophysical sources the neutrino emission can be characterized or
not by the simultaneous emission of UHE charged particles (thin or
thick sources). The situation is much more simple for top-down models
where the UHE particles are produced by the decay of massive relic
particles, remnants of the Big Bang, and still present with some
abundance in the cosmos. In this case, these relics decay into
particle-antiparticle pairs and among the decay channels the ones
producing neutrinos and photons are favoured. Since $\nu$'s only
interact weakly, once produced they can travel in the cosmos without
significant energy loss and deflection. This is completely different
from UHE charged particles or photons, which suffer an exponential
flux reduction due to the Greisen-Zatsepin-Kuzmin (GZK) cut-off
\cite{gzk} and a variation of the arrival directions in presence of
large diffused magnetic fields. Thus, even at this energy, one could
observe a significant neutrino flux and do a reasonable neutrino
astronomy.

In this framework, the new generation of cosmic ray surface
detectors, like the Pierre Auger Observatory \cite{auger}, and the
neutrino telescopes, like Amanda \cite{amanda}, Icecube
\cite{icecube}, Antares \cite{antares}, and EUSO \cite{euso} will
be able to study neutrinos of astrophysical origin in a wide range
of energies. This will improve our understanding of the
astrophysical acceleration mechanisms for bottom-up scenarios or,
if top-down models were confirmed, will give new insights on
fundamental interactions.

In the same way as the more usual components of cosmic radiation,
also UHE $\nu$'s can initiate EAS which could be detectable both
by large surface and fluorescence detectors. Leaving apart
vertical air showers, where the probability of a $\nu$ interaction
with the $\sim 1000$ g/cm$^2$ of atmospheric depth is negligibly
small, the best choice to look for a clear signature of neutrino
induced events, studying EAS in array like AUGER, is the case of
almost horizontal, skimming, or up-going air showers. In
particular, neutrino air showers at large zenith angles, contrary
to the proton ones \cite{avezas}, can be initiated deep in the
atmosphere, producing both fluorescence yield in their
longitudinal development and arriving to the ground detectors with
a non attenuated electromagnetic component. These qualitative
features of neutrino induced showers should however be supported
by simulations, which are a necessary quantitative tool for the
interpretation of the experimental data.

Neutrino detection in Auger has been previously addressed
\cite{augernu,french}, and some solution has been adopted \cite{french}
for the simulation of $\nu$ induced showers. In this respect, one has to
take into account that none of the official Monte Carlo's used by the
collaboration for shower simulation in atmosphere, CORSIKA \cite{corsika}
and AIRES \cite{aires}, treats neutrinos as primary particles. Therefore,
the authors in \cite{french} used AIRES for simulating the interaction of
muon neutrinos, injecting at low altitudes protons with a fraction of the
energy of the initial neutrinos, accompanied by a photon shower in the
case of an electron neutrino. Our approach in this framework is, instead,
to extend the capabilities of a tool like CORSIKA, already used by a large
number of people in the cosmic ray community, with the inclusion of
neutrinos to the (already long) list of primary particles which it can
handle. In case of primary neutrinos, we make a call to an existing Monte
Carlo, HERWIG \cite{herwig}, to treat only the neutrino first interaction,
and then leave in the hands of CORSIKA the products of such interaction.
The HERWIG event generator is continuously updated by particle physics
community, which makes it an extremely reliable tool for the description
of particle interactions. This represents the main advantage of our
approach.

The paper is organized as follows: in sections \ref{corsika} and
\ref{herwig} we describe the two Monte Carlo codes CORSIKA and HERWIG,
respectively. The modified version of the shower generator is outlined in
section \ref{corher}, whereas in section \ref{results} we report the
results of our simulation and in section \ref{conclusions} give our
conclusions.

\section {\bf CORSIKA Monte Carlo}
\label{corsika}

CORSIKA (COsmic Ray SImulations for KAscade) \cite{corsika} is a detailed
Monte Carlo program to study the EAS in the atmosphere initiated by
photons, protons, nuclei and many other particles. It was originally
written to perform simulations for the Kascade experiment \cite{kascade},
but during the time it became a tool which is used by many groups and its
applications range from Cherenkov telescope experiments ($E_0 \sim
10^{12}$ eV) up to giant cosmic ray surface experiments at the highest
energies observed ($E_0 > 10^{20}$ eV).

The most serious problem of EAS simulation programs is the
extrapolation of the hadronic interaction to higher energies and
into rapidity ranges which are not covered by experimental data.
In particular, the extreme forward direction is not accessible by
present collider experiments, but the particles in this
kinematical region play the more important role in the development
of EAS. In fact, they are the most energetic secondary particles
which carry deep in the atmosphere the largest energy fraction of
each collision. Therefore one has to rely on extrapolations based
on theoretical models.

The simplest hadronic model, HDPM, was produced in the 1989 by
Capdevielle \cite{capdevielle} and inspired by the Dual Parton
Model \cite{capella}. It describes the hadronic interactions of
protons at high energies in good agreement with the measured
collider data. As an alternative to the phenomenological HDPM
generator one can use other hadronic interaction models. VENUS
\cite{werner}, QGSJET \cite{qgsjet}, and DPMJET \cite{ranft}
describe the inelastic hadronic interaction in the theoretically
well founded Gribov-Regge \cite{gribregg} theory of multi-Pomeron
exchange, which has been successfully used over decades for
treating elastic and inelastic scattering of hadrons. SIBYLL
\cite{sibyll} instead is a minijet model that describes the rise
of cross-section with energy by increasing the pairwise minijet
production. Particle tracking with ionization and radiation
losses, multiple scattering and decay of unstable particles are
performed for all models in the same way. In the last version of
CORSIKA the new model NEXUS \cite{nexus} has been added, which
combines algorithms of VENUS and QGSJET with ideas based on the
data from the experiments H1 and ZEUS, and is best suited for
extrapolations up to higher energies. Further, the GHEISHA
routines \cite{fesefeldt} have been introduced to have a more
sophisticated treatment of low energy hadronic interactions with
respect to the old ISOBAR model. In fact, the hadronic models are
only used for reactions above E$_{lab}$ = 80 GeV/N, and below this
threshold the GHEISHA code is used or the recently added UrQMD
program \cite{urqmd}, designed to treat low energy hadron-nucleus
and nucleus-nucleus interactions.

In contrast to the hadronic particle production, the electromagnetic
interactions of shower particles can be calculated very precisely from
Quantum Electrodynamics. Therefore electromagnetic interactions are not a
major source of systematic errors in the shower simulation. Very well
tested packages exist to simulate these reactions in great detail, like
EGS4. CORSIKA uses an adapted version of EGS4 for the detailed simulation
of electromagnetic interactions, which includes the
Landau-Pomeranchuk-Migdal \cite{lpm} effect, the production of muon pairs
and hadrons by photons, muon bremsstrahlung and e$^+$ + e$^-$ pair
production by muons. EGS was also modified to accommodate the changes in
the atmospheric density and to compute particle production with double
precision, following the particles up to energies of typically 100 keV. In
addition, the total energy deposited along the shower axis is recorded,
all two and three body decays, with branching ratios down to 1\%, are
modelled in a kinematically correct way, and particle tracking and
multiple scattering are realized in great detail. An alternative way of
treating the electromagnetic component is to use the improved and adapted
form of the analytical NKG formula \cite{nkg} for each electron or photon
produced in the hadronic cascade. Last but not least, to account for
seasonal and geographical variations CORSIKA allows the choice of a
variety of atmospheric density profiles and the definition of new ones.

The CORSIKA program recognizes more than 50 elementary particles:
\begin{itemize}
\item $\gamma ,e^{{\pm} },\mu ^{{\pm} }$;
\item $\pi ^{o}, \pi ^{{\pm} },K^{{\pm} }$,
$K_{S/L}^{o}$ and $\eta$;
\item the baryons $p,n,\Lambda ,\Sigma ^{{\pm} },\Sigma ^{o}$, $\Xi
^{o},\Xi ^{-},\Omega ^{-}$ (and the corresponding antibaryons);
\item the resonance states $\rho ^{{\pm} }$, $\rho
^{o},K^{*{\pm} },K^{*o}$, $\overline{K}^{*o},\Delta ^{++}$, $\Delta
^{{\pm} },\Delta ^{o}$ (and the corresponding antibaryonic
resonances);
\item nuclei up to $A\leq 56$.
\end{itemize}

Concerning neutrinos, the $\nu _{e}$'s and $\nu _{\mu }$'s and the
corresponding antineutrinos produced in $\pi ,K$ and $\mu$ decays, may be
generated explicitly by using a particular option of CORSIKA, but they
cannot be chosen as primary particles. However, as discussed in the
introduction, atmospheric showers induced by neutrinos arriving to earth
are of great interest from different points of view. Here, it is worth
pointing that neutrino fluxes, at the origin expected with the ratios
$\nu_e : \nu_\mu : \nu_\tau = 1 : 2 : 0$, would evolve towards the ratios
1 : 1 : 1 due to oscillations. Therefore, the inclusion of all neutrino
flavours in the list of primaries of CORSIKA could be very helpful to the
aim of exploring a large class of phenomena in cosmic ray physics.

\section {\bf HERWIG Monte Carlo}
\label{herwig}

HERWIG \cite{herwig} is an event generator for high-energy
processes particularly suited for detailed simulation of QCD
parton showers. It provides simulation of hard lepton-lepton,
lepton-hadron and hadron-hadron scattering and soft hadron-hadron
collisions within a single package.

The main features of HERWIG are:
\begin{itemize}
\item use of angular ordering to account for initial and final-state jet
evolution in QCD with soft gluon interference;
\item color coherence of (initial and final) partons in all hard
subprocesses, including the production and decay of heavy quarks and
supersymmetric particles;
\item azimuthal correlations within and between jets due to gluon
interference and polarization;
\item jet hadronization via cluster formation based on non-perturbative gluon
splitting, and a similar cluster model for soft and underlying hadronic
events;
\item a space-time picture of event development, from parton showers to
hadronic decays, with an optional colour rearrangement model based on
space-time structure.
\end{itemize}

The main features of a hard process (with high momentum transfer)
simulated by HERWIG can be grouped in the following four classes,
presented in order of increasing scales of distance and time:
\begin{enumerate}
\item Elementary hard subprocess. Two beam particles, or their
constituents, interact and produce one or more outgoing fundamental
objects. The boundary conditions for the initial and final-state parton
showers are chosen by using the hard scale of the momentum transfer, $Q$,
and the color flow of the subprocess.
\item Initial and final-state parton showers. A parton constituent
of an incident hadron with low space-like virtuality can radiate time-like
partons. On a similar time scale, an outgoing parton with large time-like
virtuality can generate a shower of partons with lower virtuality.
Moreover, partons from the initial-state emission can produce parton
shower. The actual amount of emission is again controlled by the momentum
transfer scale of the hard subprocess, $Q$.
\item Heavy object decay. Massive object (top quarks, higgs
bosons, etc.) can decay on time scales shorter or comparable to that of
the QCD parton showers. Depending on their decay mode, they can also
produce parton showers before or after decaying.
\item Hadronization process. As a last step, to produce a realistic
simulation one has to combine the partons into hadrons. This hadronization
process takes place at low momentum transfer scale, where $\alpha_s$ is
large and perturbation theory is not applicable. Nevertheless, these kind
of processes can be described by phenomenological models. In HERWIG this
is made by terminating showering at a low scale, $Q_0 < 1$ GeV, where
colour neutral clusters are formed, which decay into the observed hadrons.
Initial-state partons are incorporated into the incoming hadron through a
soft non perturbative ``forced-branching'' phase of space-like showering.
Instead, constituent spectator partons participate in a soft ``underlying
event'' interaction, modelled on soft minimum bias hadron-hadron
collisions.
\end{enumerate}

\begin{figure}[p]
\begin{center}
\bt{l}
\epsfig{file=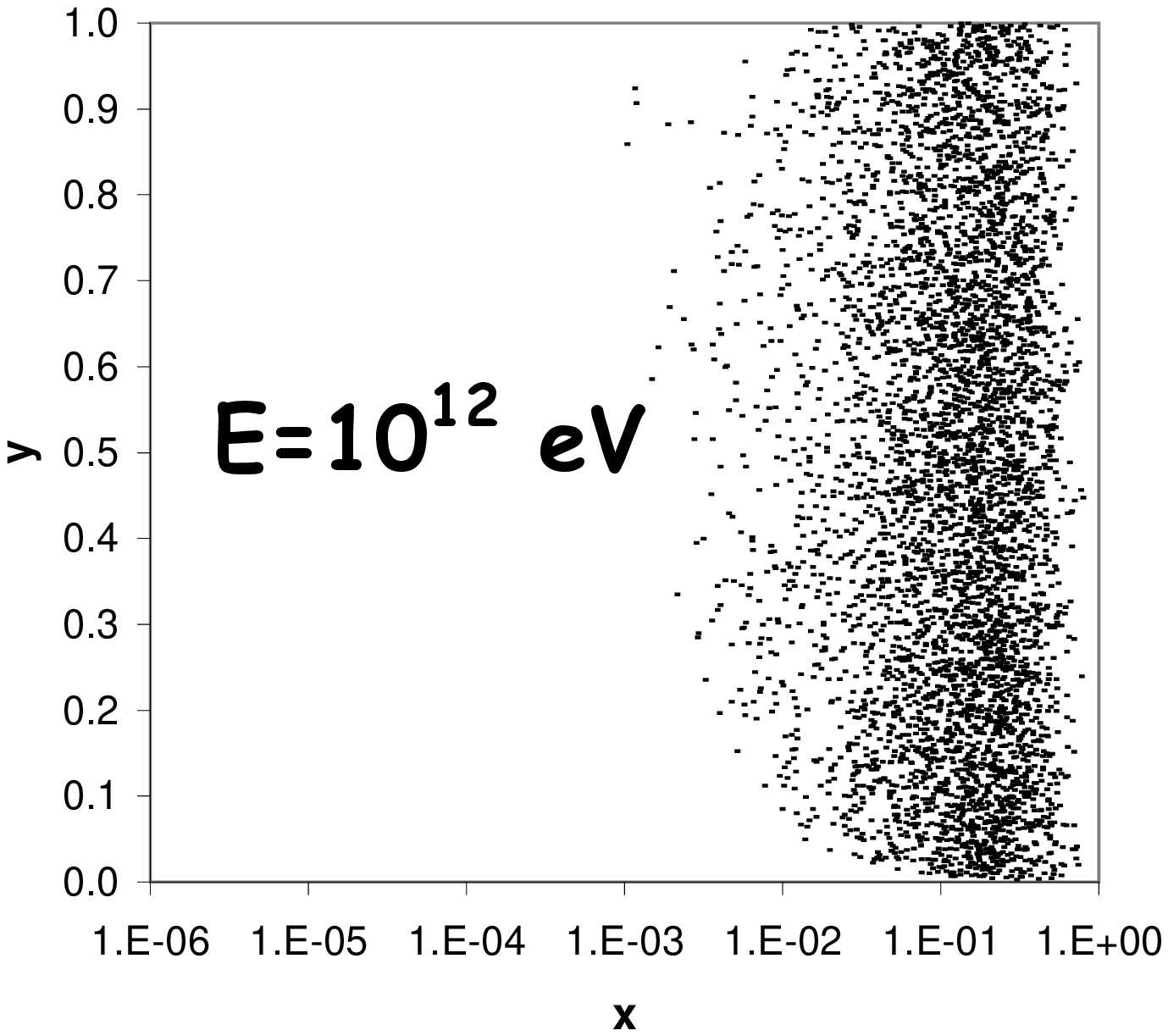,width=7.5truecm}
\\
\epsfig{file=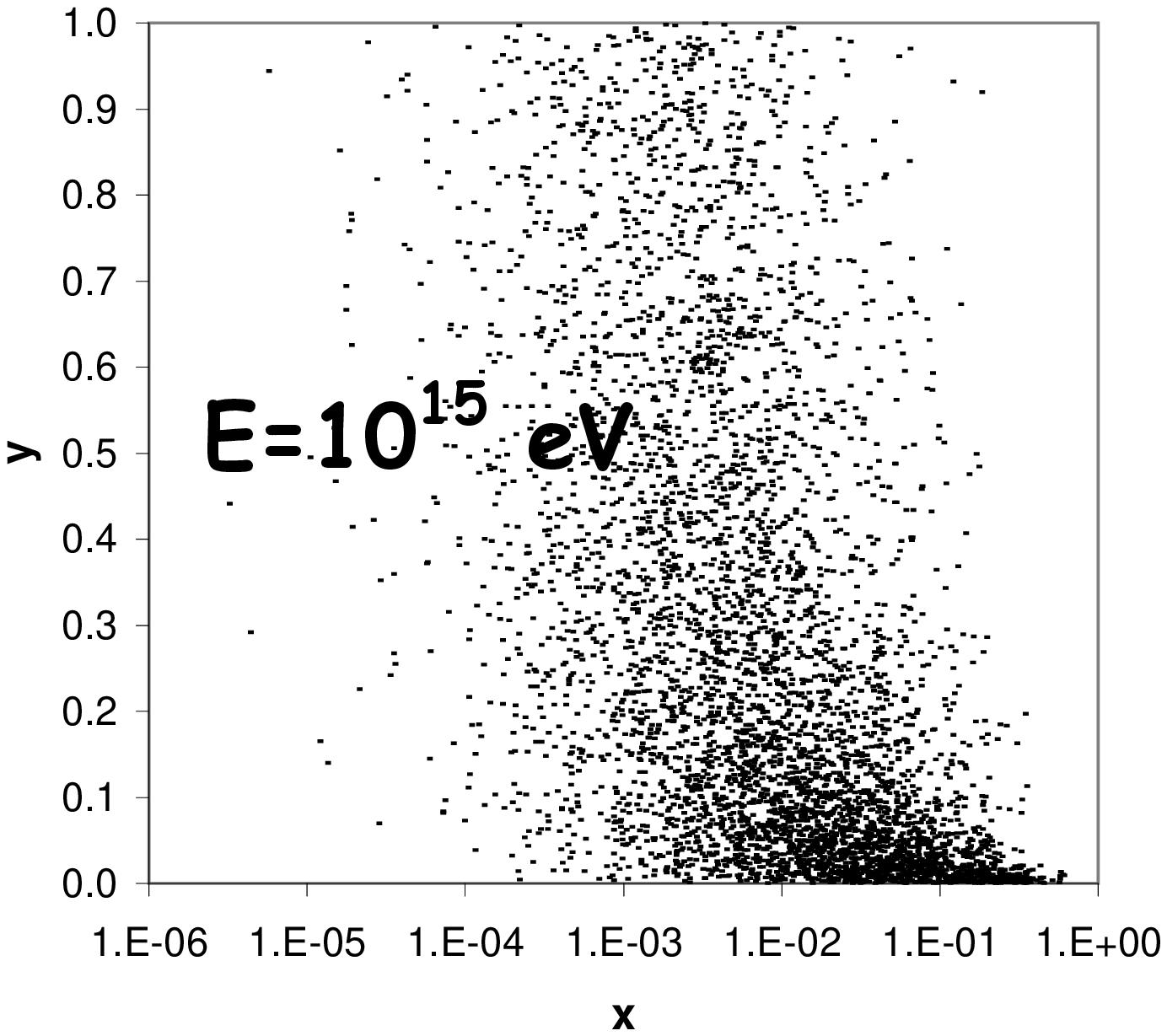,width=7.5truecm}
\\
\epsfig{file=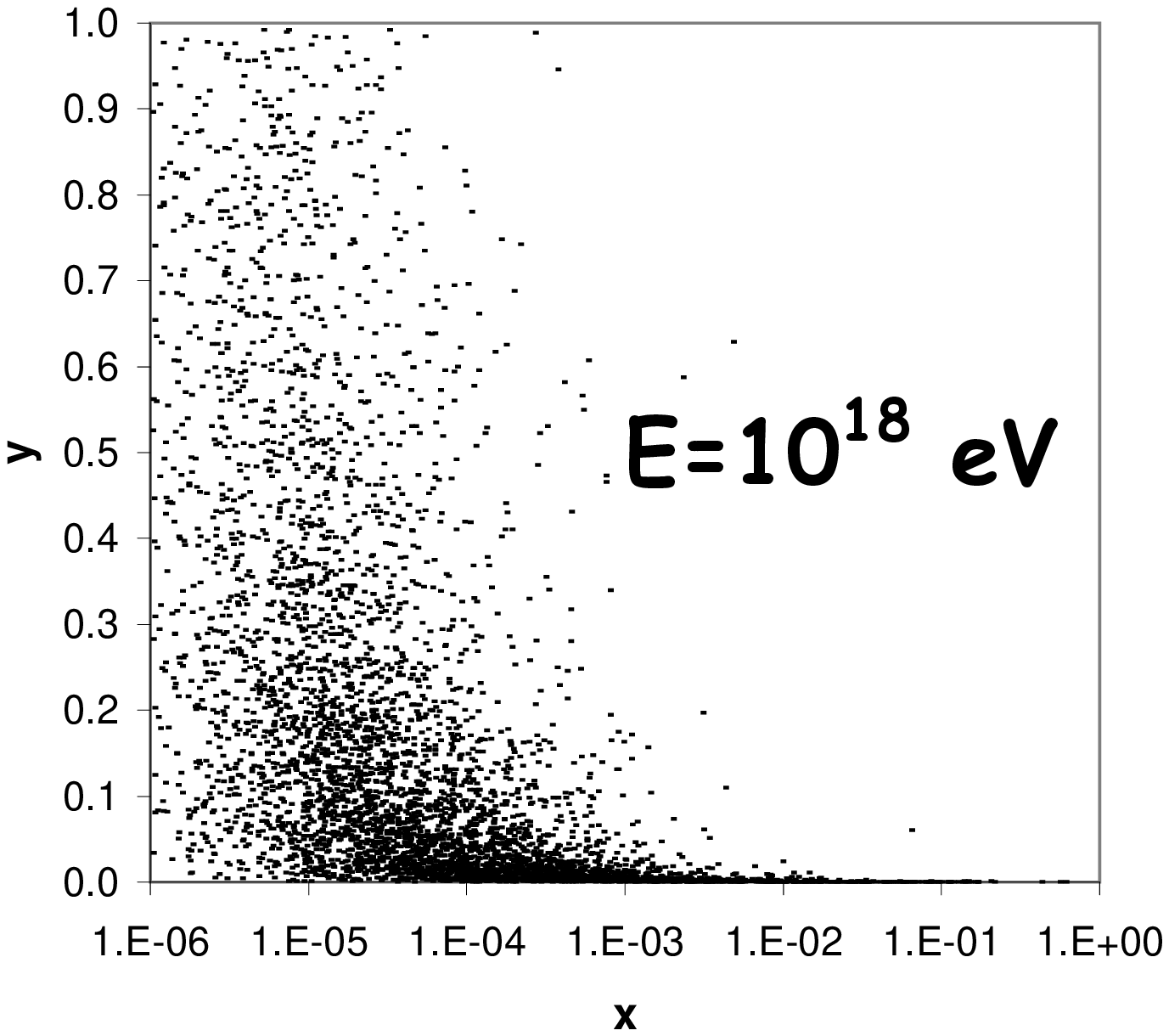,width=7.5truecm}
\et
\end{center}
\caption{Distribution in the space of the Bjorken invariant
variables $x$ and $y$, of 5000 $\nu_\mu + p \rightarrow \mu^- + X$
events. From top to bottom the neutrino energy is $10^{12}$,
$10^{15}$, and $10^{18}$ eV.}
\label{events}
\end{figure}

The version 6.5 of HERWIG is presently available and is planned to be the
last release in Fortran. Future developments, to be released in 2003, will
be implemented using C++ \cite{herwig++}.

\begin{figure}
\begin{center}
\bt{l}
\epsfig{file=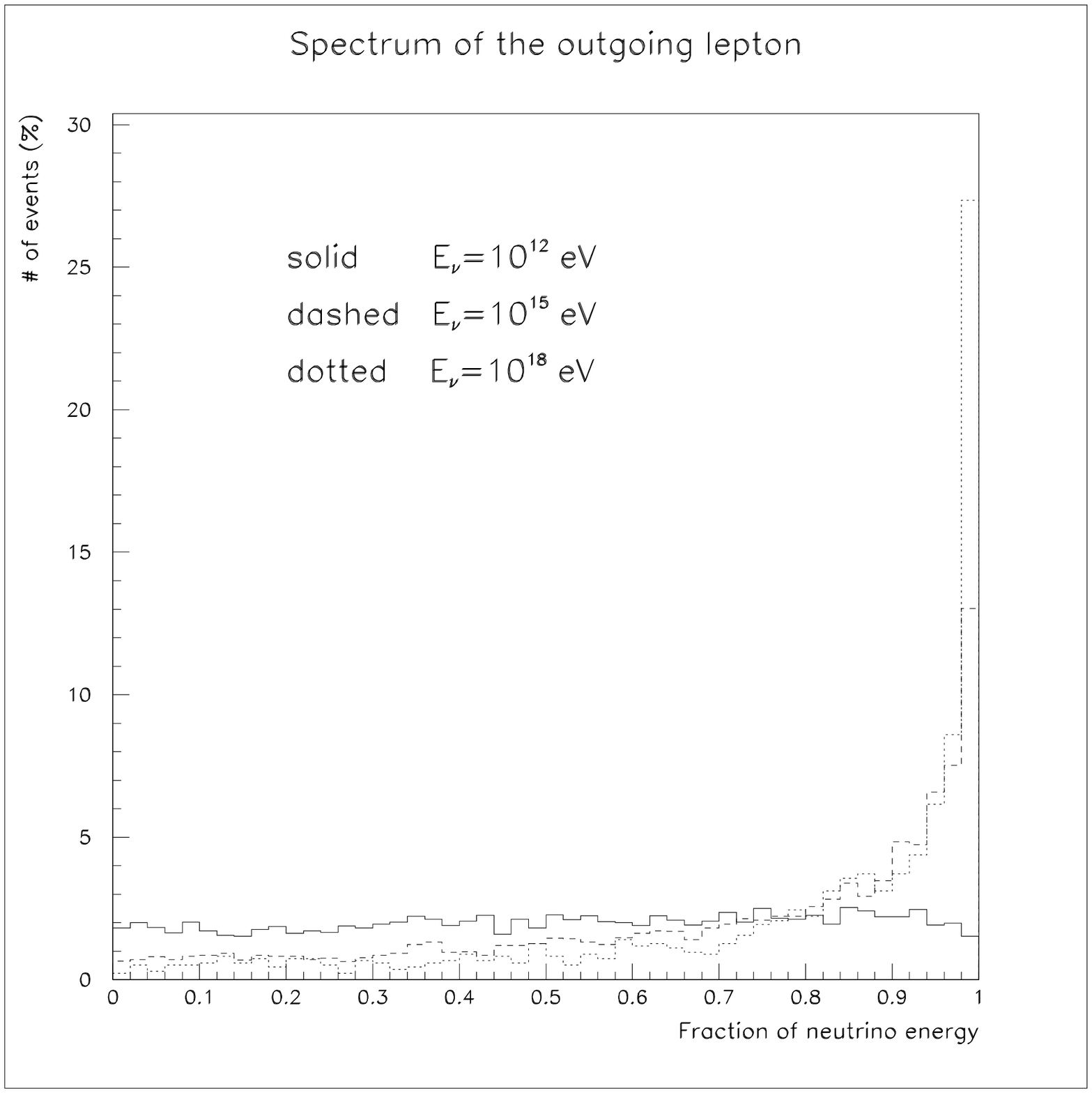,width=11truecm}
\\
\epsfig{file=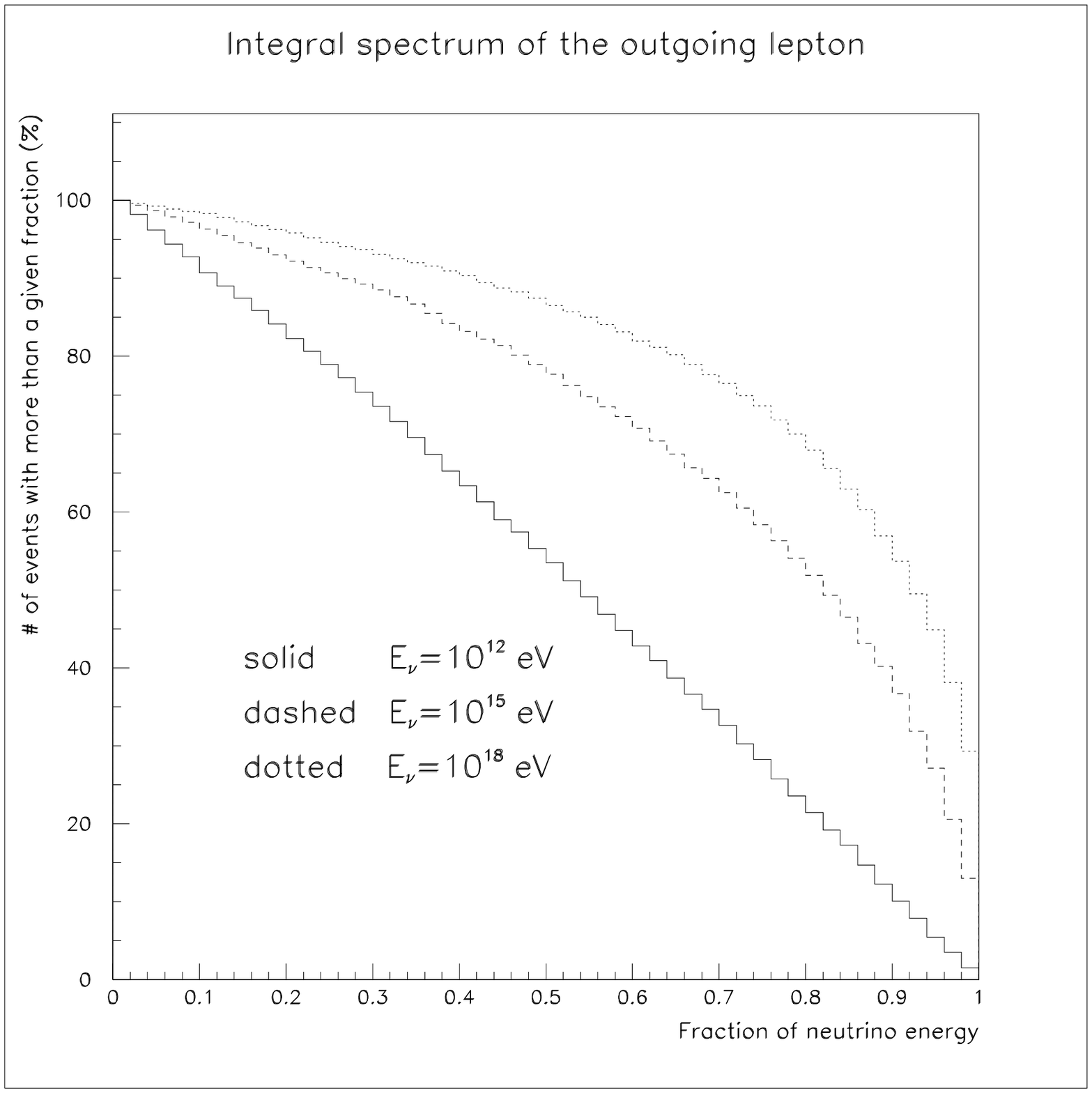,width=11truecm}
\et
\end{center}
\caption{Spectrum of the outgoing lepton for the same events
showed in Fig.\ \ref{events}. Upper (lower) plot shows the number
of events with a given fraction (the number of events with more
than a given fraction) of the neutrino energy to the outgoing
lepton.}
\label{herwspect}
\end{figure}

For the present analysis we adopted  HERWIG 6.4, which uses
standard Fortran 77. With this code a preliminary study of the
first interaction of a neutrino on a nucleon $\nu(k) \, + \, N(p)
\, \rightarrow \, l/\nu(k') \, + \, X(p')$ was performed. As far
as the structure of events in phase space is concerned, in Fig.\
\ref{events} we show, for a $\nu_\mu$ Charge Current (CC)
interaction at $E_{\nu_\mu} = 10^{12},~ 10^{15}$, and $10^{18}$
eV, the distribution of events in the space of the Bjorken
variables
\bea
x & \equiv & \frac{-q^2}{2~ p{\cdot} q} \equiv \frac{Q^2}{2~ p{\cdot} q}\, , \\
y & \equiv & \frac{p{\cdot} q}{p{\cdot} k}\, .
\eea
In the above equations $q=k-k'$ is the momentum transfer to the
nucleon. The $x$ variable represents the fraction of the parent
nucleon momentum carried by the incoming parton, while in the
nucleon rest frame $y=1-E_l/E_\nu$ is connected to the fraction of
neutrino energy to the outgoing lepton. Fig.\ \ref{events} shows
that neutrinos with higher energies probe the Parton Distribution
Functions (PDF's) inside the nucleon at lower $x$. At the same
time, the rise of $E_\nu$ results in a larger fraction of the
primary energy to the outgoing lepton, $E_l/E_\nu=1-y$. This is
better seen in Fig.\ \ref{herwspect} (upper plot), where one can
see the spectrum of the outgoing lepton resulting from the same
events shown in Fig.\ \ref{events}. Lower plot in Fig.\
\ref{herwspect} shows that the outgoing lepton produced in the
first interaction of the neutrino carries away more than 90\% of
the neutrino energy in 10\% of the cases at $E_\nu = 10^{12}$ eV,
but this percentage increases to 37\% at $E_\nu = 10^{15}$ eV and
to 54\% at $E_\nu = 10^{18}$ eV. Note that the above features of
Fig.\ \ref{events} and Fig.\ \ref{herwspect} are encoded in the
particular dependence of the $\nu-N$ cross-section on $x$ and $y$.

\begin{figure}
\begin{center}
\bt{l}
\epsfig{file=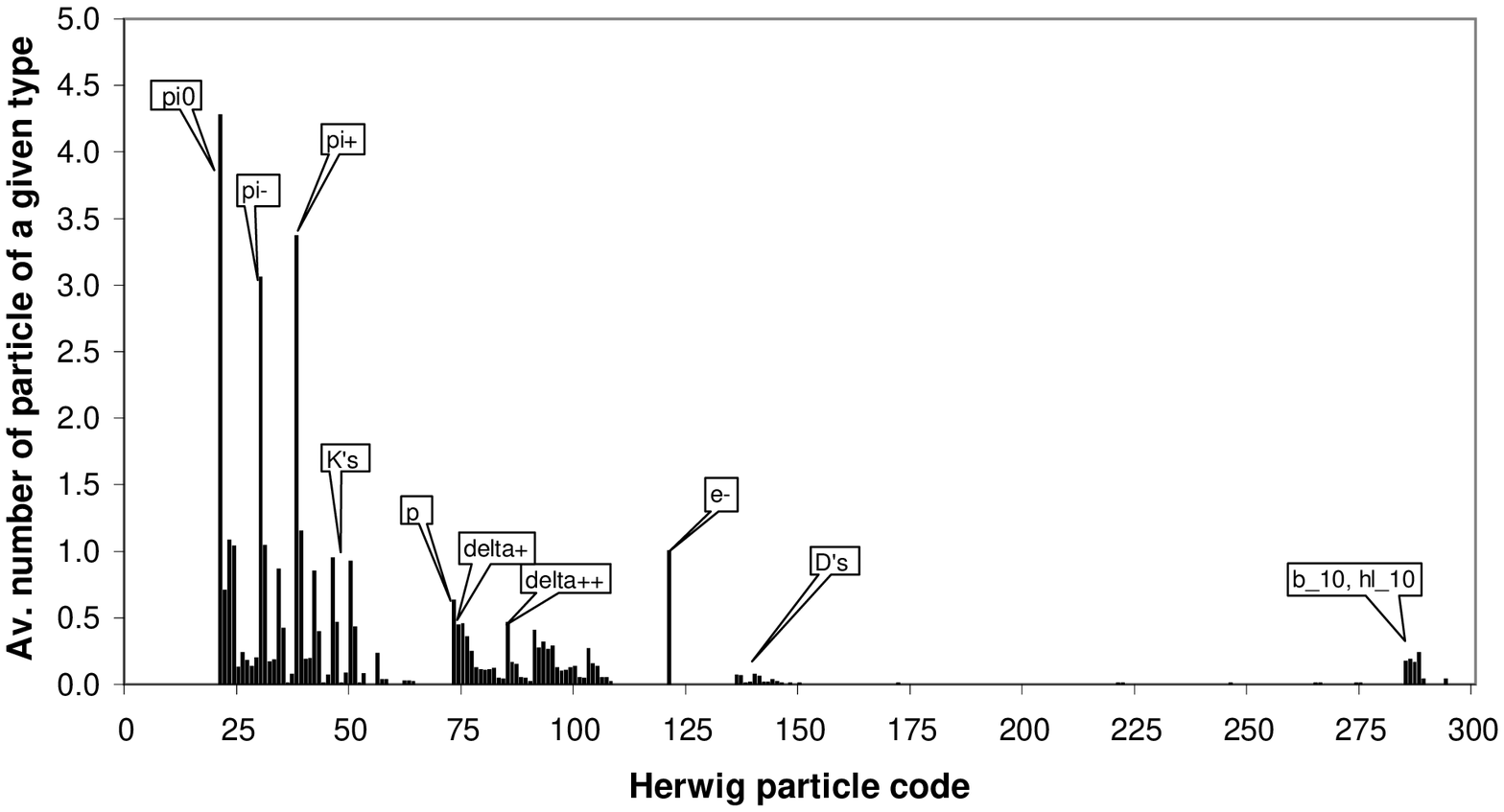,width=16truecm}
\\
\epsfig{file=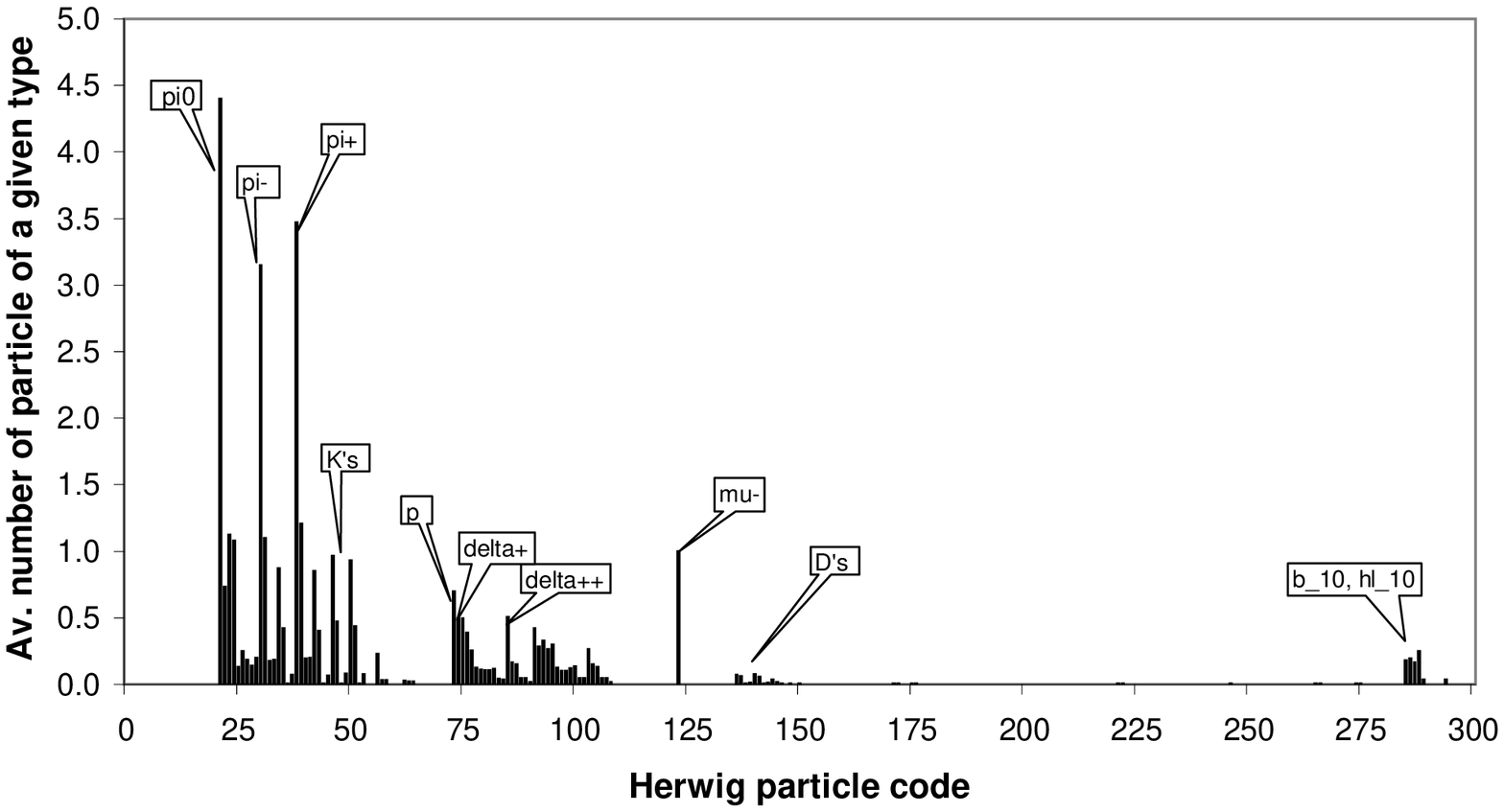,width=16truecm}
\et
\end{center}
\caption{Average numbers of different particle types produced in
the processes $\nu_e + p \rightarrow e^- + X$ (upper plot) and
$\nu_\mu + p \rightarrow \mu^- + X$ (lower plot) at $E_\nu =
10^{15}$ eV.}
\label{FIpart}
\end{figure}

It is also worth noticing that the default PDF's used by HERWIG,
the MRST Leading Order PDF's \cite{mrst98}, are valid in the
ranges $x \geq 10^{-6}$ and $1.25$ GeV$^2 \leq Q^2 \leq 10^{7}$
GeV$^2$. This means that one needs extrapolations of the PDF's
only for energies of the incoming neutrino $\gsim 10^{18}$ eV,
which are much larger than the values we have considered in this
analysis.

Last but not least, we have studied the typical particle
population of an event simulated by HERWIG. At this aim we have
selected, from the output of the Monte Carlo, the produced First
Interaction (FI) particles.  The average number of a given
particle type is shown in Fig.\ \ref{FIpart} for $\nu_e$ and
$\nu_\mu$ at $E_{\nu} = 10^{15}$ eV. From the plots it is clear
that, together with ordinary particles like $\pi$'s, K's, N's, and
$\Delta$'s, some other species are produced which are not yet
recognized by CORSIKA (like charmed particles). However, this
problem can be overcome at the energies of interest for the
present analysis, as we will see in the next section.

\section{A modified version of CORSIKA for neutrino simulation}
\label{corher}

For the present work we used the CORSIKA version 6.014, making our changes
on the Fortran file extracted from the {\it .car} source by using the
software CMZ \cite{cmz}. In the extraction, we used HDPM as the hadronic
interaction model, EGS4 for describing the electromagnetic interactions,
and the CURVED atmosphere option. The choice of HDPM is due to the fact
that this model is the simplest among the hadronic ones available in
CORSIKA, even if it agrees with experimental data only up to the energies
of $\sim 10^{16}$ eV. This motivated our choice of $E_{primary} = 10^{15}$
eV for the simulations on which the present analysis is based.

\begin{figure}
\begin{center}
\epsfig{file=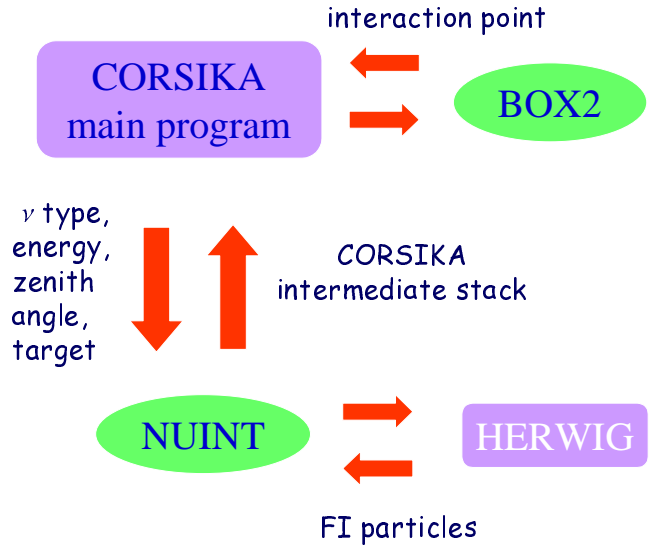,width=10truecm}
\end{center}
\caption{Flow diagram of the neutrino modified version of CORSIKA.}
\label{flow}
\end{figure}

The implementation of the neutrino modified version of CORSIKA consists of
two main steps, as showed in the flow diagram in Fig.\ \ref{flow}. The
first one is a modification in the subroutine BOX2, where the interaction
point of any particle is calculated, and the second one is a change in the
main program, for generating the first interaction of the primary
particle.

Actually, the first modification was made only for completeness, since the
cross-section of neutrino interaction with matter is so small that, even
in the case of the largest zenith angles of the primary neutrino, meaning
a crossed column depth of $\sim 36000$ g/cm$^2$, the probability of
producing a shower is very small. Due to this, we always asked CORSIKA to
make the neutrino first interaction at a fixed height. In any case, in
view of future implementations of the neutrino first interaction to higher
energies, in BOX2 the neutrino cross-section quoted in \cite{gqrs} can be
used, that is
\bea
\sigma_{tot} (\nu N) &=& 7.84~ 10^{-36}~{\rm cm}^2~ \left(
\frac{E_\nu}{1~{\rm GeV}} \right)^{0.363}~, \\
\sigma_{tot} (\bar\nu N) &=& 7.80~ 10^{-36}~{\rm cm}^2~ \left(
\frac{E_\nu}{1~{\rm GeV}} \right)^{0.363}~,
\eea
which have 10\% accuracy within the energy range $10^{16}~{\rm eV} \leq
E_\nu \leq 10^{21}~{\rm eV}$.

As far as the second change is concerned, we inserted in the main
program a conditional call to a link subroutine, NUINT, executed
only for a primary neutrino and in the first interaction case. In
this subroutine, the necessary quantities for making the call to
the HERWIG event generator are generated from the input
quantities, that is the neutrino type\footnote{Tau neutrino were
not considered in this analysis since their charged current
interaction would have produced a tau lepton which, in the 6.014
version of CORSIKA, is not handled by the program. On the other
side, its treatment with HERWIG would have gone beyond the aims of
this study.} ($\nu_e$, $\bar\nu_e$, $\nu_\mu$, $\bar\nu_\mu$), its
energy and zenith angle, and the target nucleon. This is chosen
randomly between a proton or a neutron, and the process, CC or
Neutral Current (NC) deep inelastic scattering, is selected by
using the comparison of the respective cross-sections \cite{gqrs},
\bea
\sigma_{CC} (\nu N) &=& 5.53~ 10^{-36}~{\rm cm}^2~ \left(
\frac{E_\nu}{1~{\rm GeV}} \right)^{0.363}~, \\
\sigma_{NC} (\nu N) &=& 2.31~ 10^{-36}~{\rm cm}^2~ \left(
\frac{E_\nu}{1~{\rm GeV}} \right)^{0.363}~, \\
\sigma_{CC} (\bar\nu N) &=& 5.52~ 10^{-36}~{\rm cm}^2~ \left(
\frac{E_\nu}{1~{\rm GeV}} \right)^{0.363}~, \\
\sigma_{NC} (\bar\nu N) &=& 2.29~ 10^{-36}~{\rm cm}^2~ \left(
\frac{E_\nu}{1~{\rm GeV}} \right)^{0.363}~.
\eea

The call to HERWIG is made in a standard way (see, for example, a
sample main program at the HERWIG home page \cite{hermain}). We
select the first {\it good} event from a total of MAXEV=100. This
means that, in the user's routine for terminal calculations (in
\cite{hermain} this routine is called HWAEND) several checks are
made before accepting the event: a) the event is discarded if the
total energy of the daughter FI particles differs by more than 1\%
from the primary neutrino energy; b) the event is discarded if the
values of the invariant variables which describe the deep
inelastic process, $x$ and $Q^2$, are out of the ranges where the
PDF's are given\footnote{Strictly speaking, this would imply a
bias against values favoured by the cross-section but out of the
present acceptable ranges of $x$ and $Q^2$. However, at the energy
we are concerned, the cross-section is dominated by $x$ and $Q^2$
values in the accepted ranges (see middle plot in Fig.\
\ref{events}).}. In this respect, we used the MRST Leading Order
PDF's \cite{mrst98}, which are the default PDF's in HERWIG.

When selecting from the output of HERWIG the FI particles, we have to
face a difficulty: the typical zoo of FI particles we obtain (see
Fig.\ \ref{FIpart}) can, and actually does in most of the cases,
include some products which are not present in the list of particles
recognized by CORSIKA. At the energy we have chosen for this analysis,
$E_\nu = 10^{15}$ eV, the average fraction of events with more than
10\% of the primary energy in these unrecognized products is $\sim
31\%$. On the other side, the $\gamma$ factor of these particles at
this energy is such that they would not cover a very long distance in
the atmosphere, and an acceptable solution to the problem is to
substitute them directly with their decay products. This is made inside
the routine HWAEND, where, once determined that the current particle
is not among the ones treated by CORSIKA, it is replaced with its
daughter particles already produced by HERWIG.

Once the FI particles have been selected, the array SECPAR of CORSIKA,
filled in the routine HWAEND with all the information about the
current particle, is used for transferring them to the CORSIKA
intermediate stack STACKINT (these particles are then transferred by
CORSIKA to the real stack, STACK, and then written to the output
file).

Besides the changes we have described, some other minor modifications
were made, concerning the data sent to the output, for collecting all
the information on the neutrino first interaction.

\section{Results}
\label{results}

With the new neutrino version of CORSIKA we generated a series of
30 showers for each neutrino species ($\nu_e$, $\bar\nu_e$,
$\nu_\mu$, $\bar\nu_\mu$), 10 for each zenith angle in the set
$\theta = 70^\circ,~ 75^\circ,~ 80^\circ$. These inclinations are
such that one has to use the option of curved atmosphere in
CORSIKA.  The energy of the primary $\nu$ was $E_\nu = 10^{15}$
eV. At the same time, we produced, with the version 6.014 of
CORSIKA, a correspondent set of proton showers with the same
inclinations and primary energy. In order to make the comparison
between the two class of showers, the neutrino first interaction
was realized at the same point of the corresponding proton shower
(note that, on the contrary, the proton first interaction is
driven by CORSIKA according to the proton-air cross-section). The
observation level was chosen in such a way that an EAS with zenith
angle $\theta = 70^\circ$ is intercepted at an atmospheric depth
of $760$ g/cm$^2$. This situation is very similar to what occurs
at the Auger level for the same particles but with a rescaled
energy larger than $10^{18}$ eV. Changing the inclination of the
primary to $\theta = 75^\circ$ and $80^\circ$ the observation
level correspondingly moves to a depth of $1005$ and $1497$
g/cm$^2$, respectively.

\begin{figure}
\begin{center}
\epsfig{file=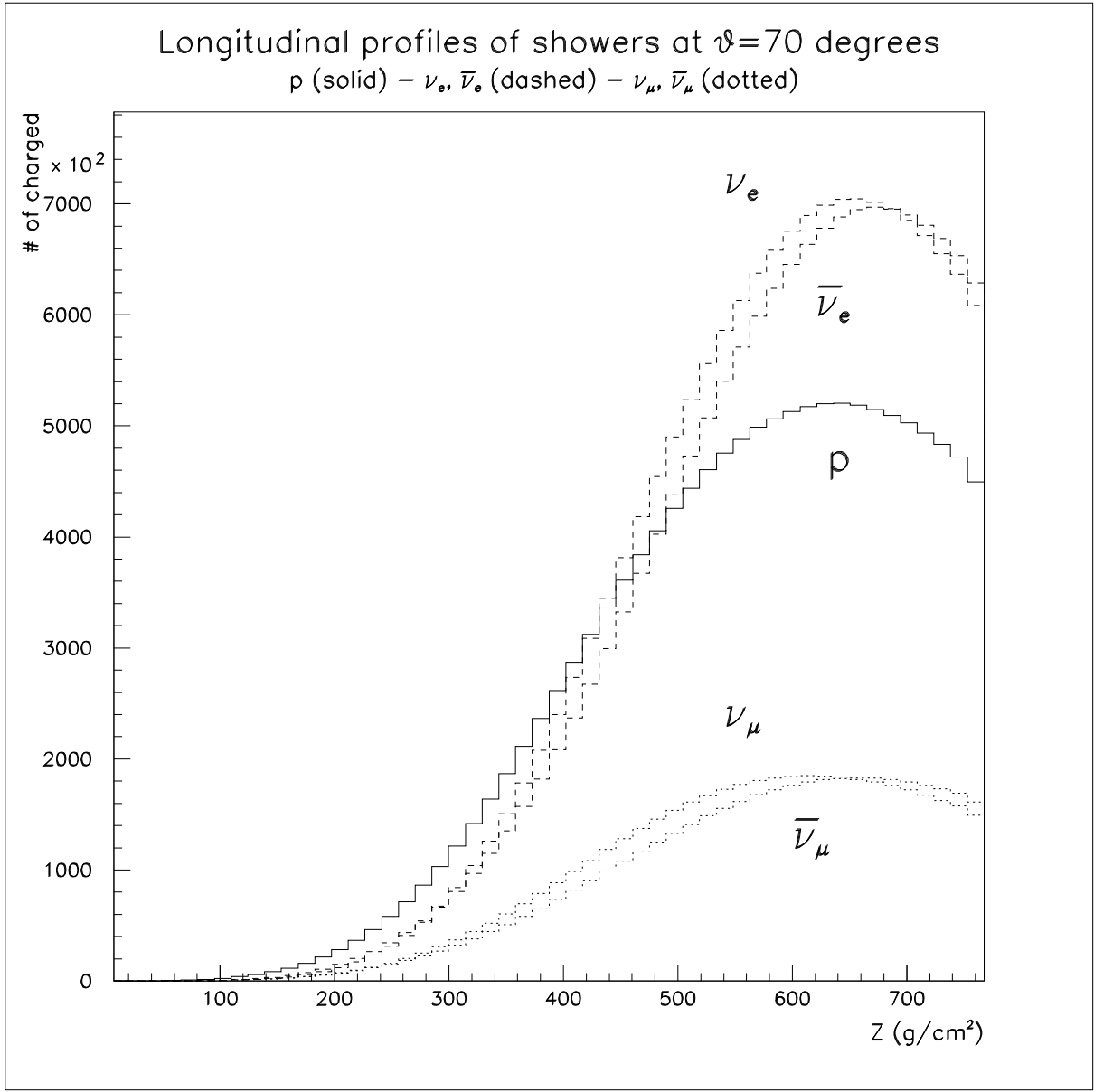,width=16truecm}
\end{center}
\caption{Average longitudinal profiles of showers induced by $p$
(solid), $\nu_e$ and $\bar\nu_e$ (dashed), $\nu_\mu$ and $\bar\nu_\mu$
(dotted) at the primary energy of $10^{15}$ eV and $\theta =
70^\circ$.}
\label{long70}
\end{figure}

\begin{figure}
\begin{center}
\epsfig{file=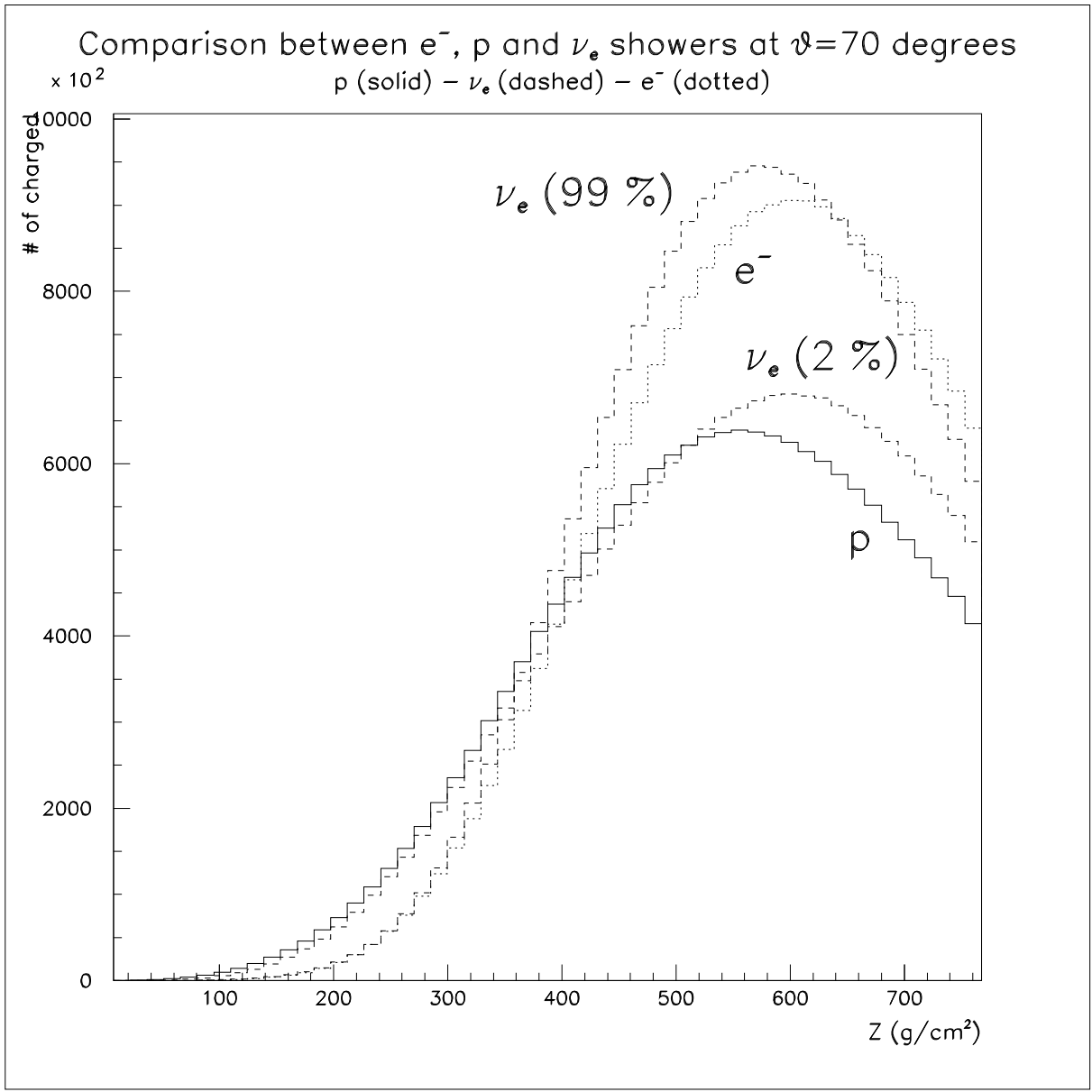,width=16truecm}
\end{center}
\caption{Comparison between the longitudinal profiles of showers
induced by $p$ (solid), $\nu_e$ (dashed), and $e^-$ (dotted) with
a primary energy of $10^{15}$ eV. The dashed curves correspond to
$\nu_e$ showers with 99\% and 2\%, respectively, of the neutrino
energy to the outgoing electron.}
\label{epnucomp}
\end{figure}

\begin{figure}
\begin{center}
\epsfig{file=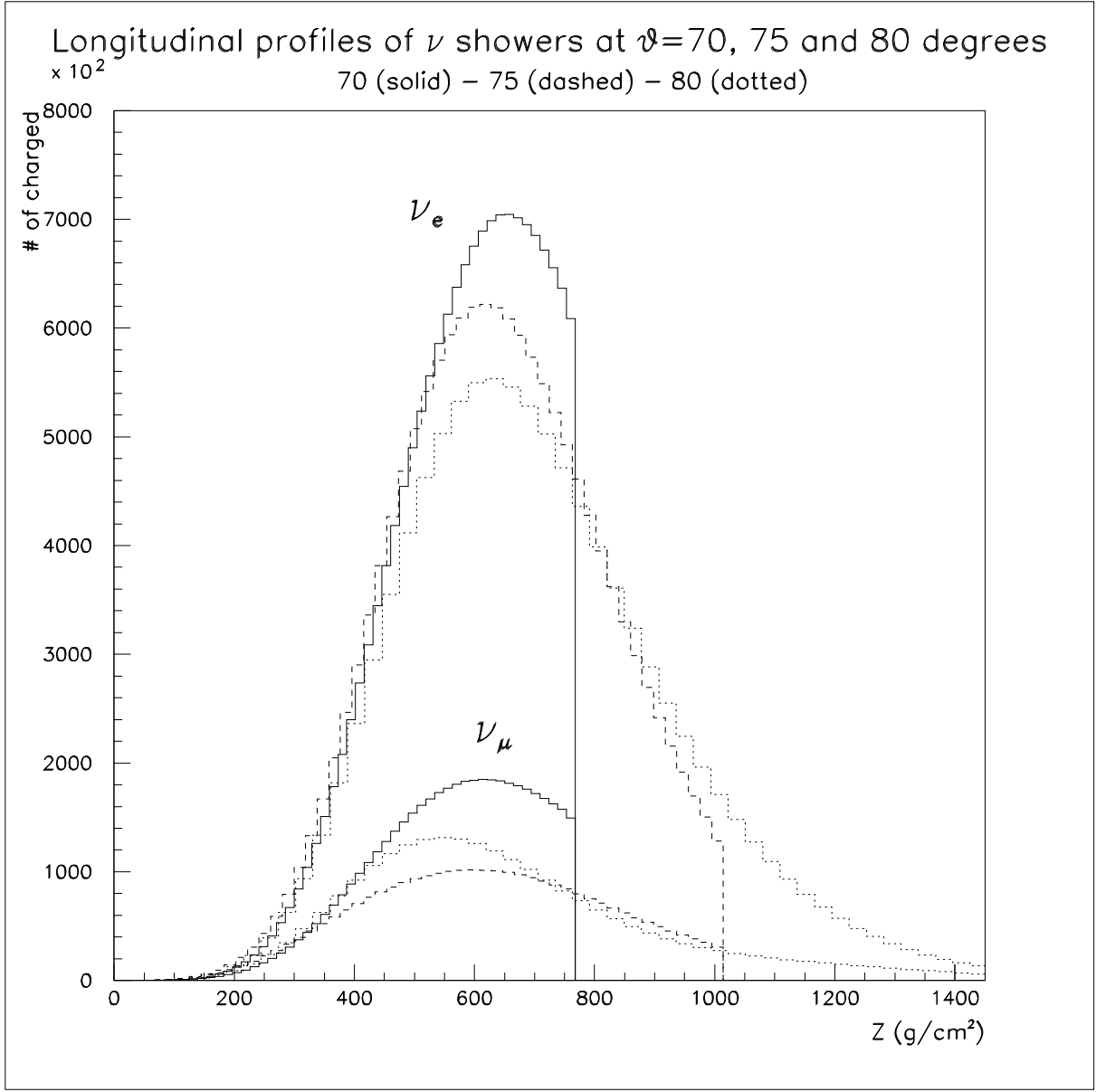,width=16truecm}
\end{center}
\caption{Average longitudinal profiles of showers induced by
$\nu_e$ (upper curves) and $\nu_\mu$ (lower curves) at the primary
energy of $10^{15}$ eV and $\theta = 70^\circ$ (solid), $75^\circ$
(dashed), and $\theta = 80^\circ$ (dotted).}
\label{longitudinal}
\end{figure}

\begin{figure}
\begin{center}
\epsfig{file=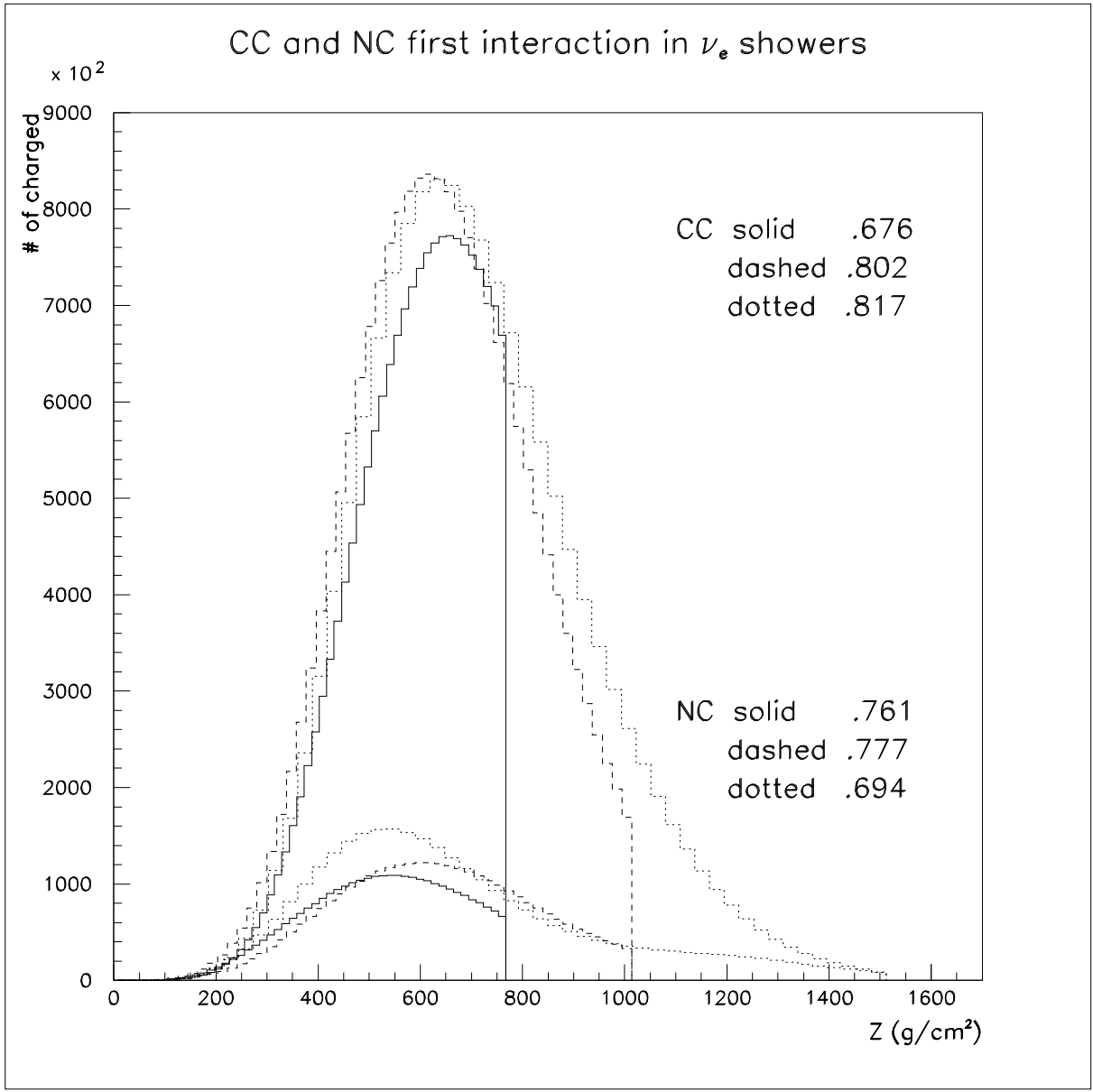,width=16truecm}
\end{center}
\caption{Average longitudinal profiles of showers induced by a
$\nu_e$ developing a CC first interaction (upper curves) or a NC
one (lower curves) at the primary energy of $10^{15}$ eV and
$\theta = 70^\circ$ (solid), $75^\circ$ (dashed), and $\theta =
80^\circ$ (dotted). The average fraction of the primary energy
delivered to the secondary lepton is reported for each curve.}
\label{ccnc}
\end{figure}

\begin{figure}
\begin{center}
\epsfig{file=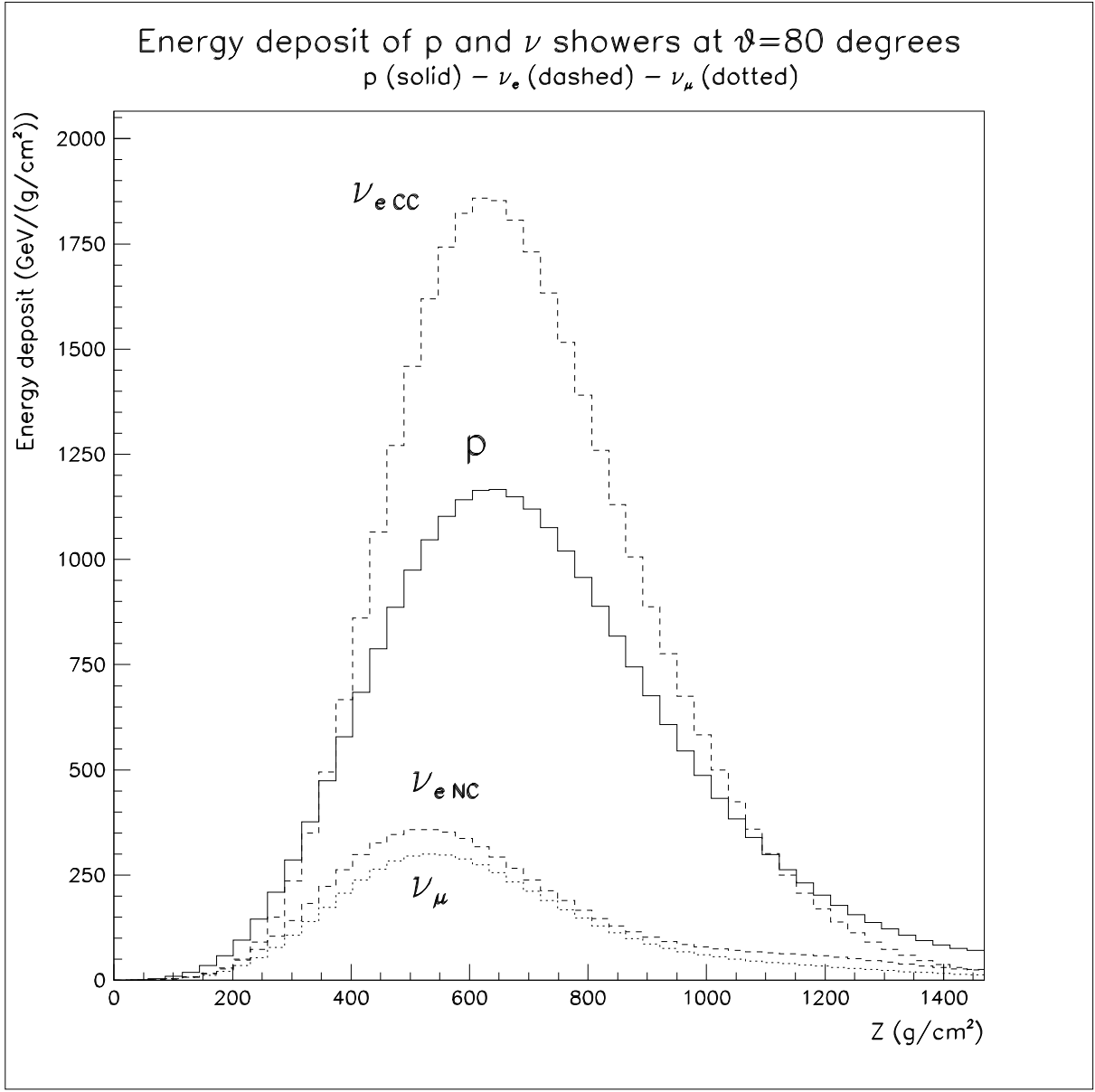,width=16truecm}
\end{center}
\caption{Comparison of the average energy deposit for showers
induced by $p$ (solid), $\nu_e$ (dashed) and $\nu_\mu$ (dotted) at
the primary energy of $10^{15}$ eV and $\theta = 80^\circ$. For
$\nu_e$ the CC and NC average components are shown separately.}
\label{edep80}
\end{figure}

\begin{figure}
\begin{center}
\epsfig{file=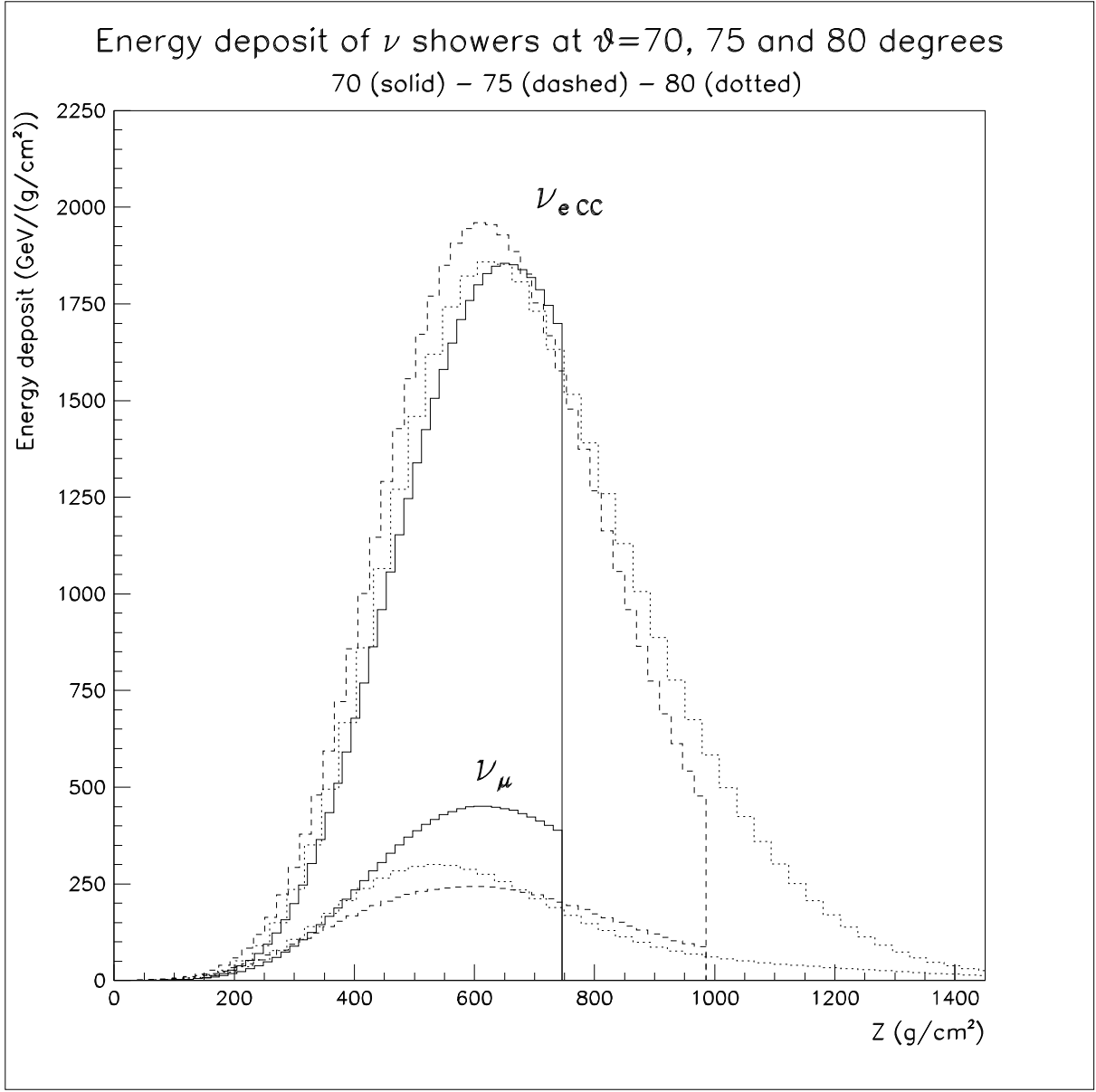,width=16truecm}
\end{center}
\caption{Comparison of the average energy deposit for showers
induced by $\nu_e$ (CC component) and $\nu_\mu$ at the primary
energy of $10^{15}$ eV and $\theta = 70^\circ$ (solid), $75^\circ$
(dashed), and $80^\circ$ (dotted).}
\label{edep}
\end{figure}

\begin{figure}
\begin{center}
\epsfig{file=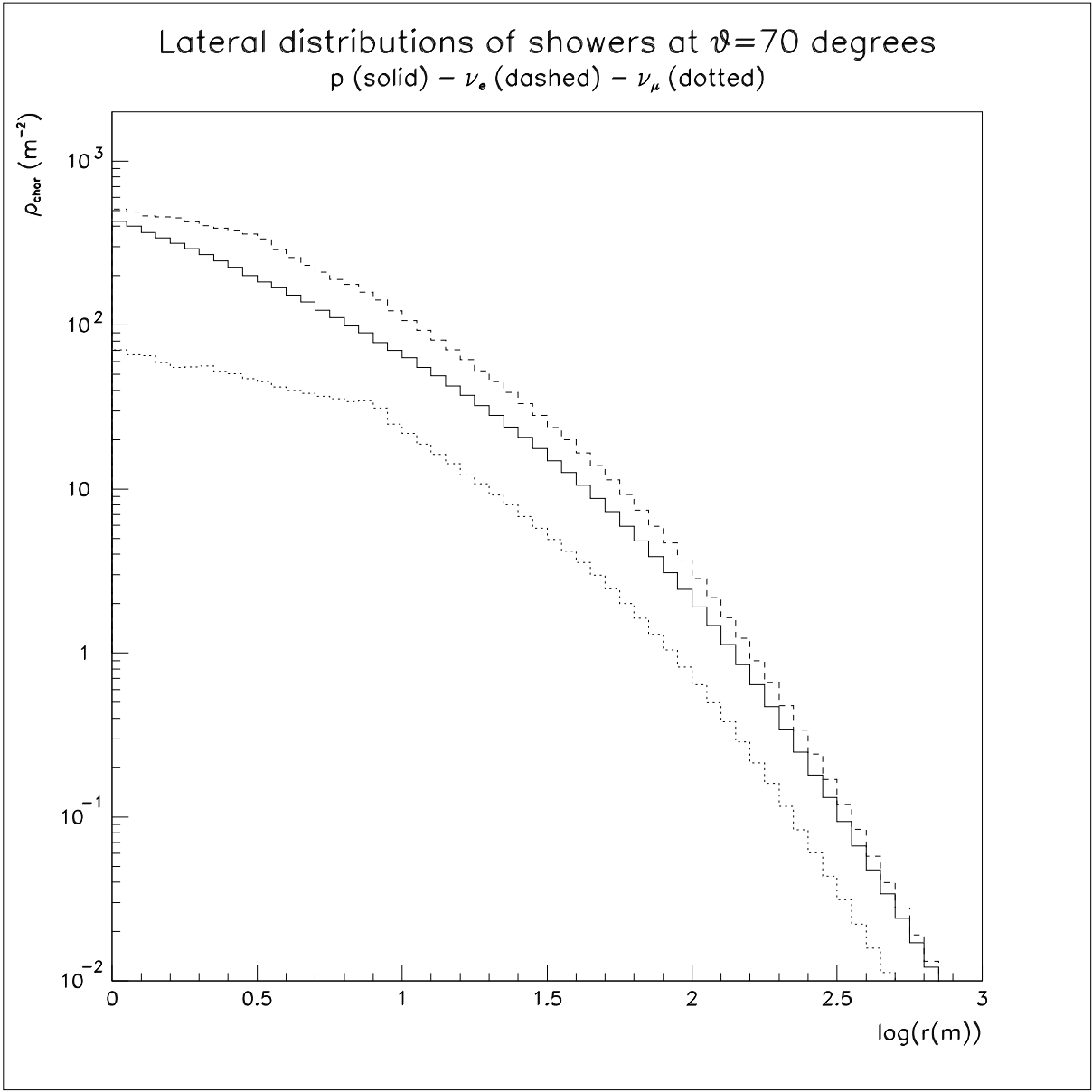,width=16truecm}
\end{center}
\caption{Average lateral density distributions of charged
particles versus the distance from the shower axis at a slant
depth of 760 g/cm$^2$ for the showers induced by $p$ (solid),
$\nu_e$ (dashed) and $\nu_\mu$ (dotted) at the primary energy of
$10^{15}$ eV and $\theta = 70^\circ$.}
\label{lateral}
\end{figure}

We analyzed each set of 10 showers at a given primary zenith angle
in order to obtain an average behaviour. The results are showed in
Figs.\ \ref{long70}-\ref{lateral}. We show in Fig.\ \ref{long70}
the comparison of the longitudinal profiles of charged particles
versus the slant atmospheric depth for proton, electron and muon
neutrinos and antineutrinos at $\theta=70^\circ$. From this plot,
no significant difference between neutrinos and antineutrinos of
the same type is appreciable. This feature results to be
independent of the shower inclination, and hence hereafter the
plots for antineutrinos will be omitted. For muon neutrinos the
particle number at the maximum is lower than for protons, while
for electron neutrinos is larger. This has a unique explanation in
terms of the outgoing lepton spectrum. In fact, for $E_\nu =
10^{15}$ eV the outgoing lepton produced in the first interaction
of the neutrino carries away more than 90\% of the neutrino energy
in 37\% of the cases (see Fig.\ \ref{herwspect}). In this respect
it is worth noticing that, for a given inclination, each set of 10
$\nu_\mu$ showers has been produced by using the same seeds of the
corresponding $\nu_e$ ones; consequently, the outgoing lepton
energy distribution is the same, as well as the percentage of CC
events with respect to NC ones\footnote{The NC interaction has a
cross-section which, in the interesting range of energies, is
roughly an order of magnitude lower than the one for a CC one.
Actually, in the set of 30 produced showers the fraction of events
NC/(NC+CC) is 8/30 (1/10 for $\theta = 70^\circ$).}. The energy of
the outgoing lepton does not contribute to the development of the
shower, and is then undetectable, either for a neutrino NC
interaction or for a muon neutrino CC one which produces a very
energetic muon hitting the ground at the core shower. On the other
side, a CC interaction of a $\nu_e$ in most of the cases produces
an electron with a large fraction of primary energy, and less
energetic hadronic products. From these first interaction
particles a mixed shower is generated, partly electromagnetic and
partly hadronic. By increasing the energy fraction delivered to
the electron, the electromagnetic features of this shower become
dominant with respect to the hadronic ones. In this case the rise
in the number of charged particles (predominantly electrons and
positrons) stops later, since the critical energy in an
electromagnetic shower is lower than in a hadronic one. This is
confirmed in Fig.\ \ref{epnucomp}, where the showers initiated by
a proton and an electron neutrino (with an energy fraction to the
secondary electron of 99\% and 2\%, respectively) are plotted for
$\theta = 70^\circ$, and compared to a corresponding $10^{15}$ eV
electron shower\footnote{This primary electron has been injected
in the atmosphere at the point where the secondary electron is
produced in the first interaction of the primary $\nu_e$.}. When
the FI electron in the $\nu_e$ shower takes 99\% of the neutrino
energy the height of the $e$ and $\nu_e$ longitudinal peaks are
comparable, while when the fraction of neutrino energy to the FI
electron is 2\% the height of the maximum in the $\nu_e$
longitudinal profile is similar to the proton one.

In Fig.\ \ref{longitudinal} the longitudinal developments at
$\theta=70^\circ$, $75^\circ$ and $80^\circ$ are shown for $\nu_e$
and $\nu_\mu$ primaries. In principle, the average longitudinal
profile of charged particles versus the slant depth should not
depend on the zenith angle. However, if the average, as in our
case, is performed on a small number of simulations one has to
take into account the Monte Carlo fluctuations. This is actually
the reason for the different features of $\nu_e$ and $\nu_\mu$
profiles at different angles reported in Fig.\ \ref{longitudinal}.
The origin of these fluctuations is mainly twofold: the event can
be CC or NC mediated and the energy of the outgoing lepton has the
distribution showed in Fig.\ \ref{herwspect}. In Fig.\ \ref{ccnc},
in order to disentangle the two effects, we plot the $\nu_e$
average showers for $\theta=70^\circ$, $75^\circ$ and $80^\circ$,
having separated the CC and NC first interaction events. The
residual difference between the curves belonging to one of the
previous categories is due to the different energy fraction of the
outgoing lepton, which is indicated in the plot. Note that, as
expected, a smaller fraction of energy to the electron in the CC
case (upper curves in Fig.\ \ref{ccnc}) results in a decrease of
the height of the maximum, while a smaller fraction to the
electron neutrino in the NC case (lower curves in Fig.\
\ref{ccnc}) amounts to an increase.

As pointed out in Ref. \cite{edep}, the charged particle energy
deposit is a quantity which is directly connected to the
fluorescence yield produced in the longitudinal development of a
shower. Actually, the authors of Ref. \cite{edep} note a good
agreement between the shapes of the energy deposit and the charged
particle profile. This is due to the ionization energy loss per
particle, which is not much dependent on the energy. As a
consequence, for slant depth $\geq 300$ g/cm$^2$, the two curves
differ for a mere multiplicative factor (see Fig.\ 7 in
\cite{edep}). The comparison between the energy deposit for
proton, $\nu_e$ and $\nu_\mu$ induced showers, at
$\theta=80^\circ$, is reported in Fig.\ \ref{edep80}. As shown in
this graph, the CC $\nu_e$ curve is much higher than the NC
$\nu_e$ and $\nu_\mu$ curves, since in both the last cases the
outgoing lepton does not produce further ionization. This implies
that, at least for a primary energy of $\sim 10^{15}$ eV, the
threshold for the detection of a CC $\nu_e$ shower by fluorescence
is lower than for a NC $\nu_e$ or a $\nu_\mu$ one. On the other
side, Fig.\ \ref{edep} shows the dependence on the zenith angle of
the energy deposit for $\nu_e$ and $\nu_\mu$ induced showers.

Finally in Fig.\ \ref{lateral} the lateral distributions of
charged particles versus the distance from the shower axis for
$p$, $\nu_e$, and $\nu_\mu$ induced showers at $\theta=70^\circ$
are plotted. The curves confirm the clear differences already
noted between the showers induced by the two neutrino flavours.

\section{Conclusions and outlook}
\label{conclusions}

In this paper we have described the results obtained by a modified
version of CORSIKA, accounting also for neutrinos as primary
particles. This tool uses a call to the HERWIG Monte Carlo to
implement the first interaction of the neutrino. The analysis of
the simulated showers for a primary energy of $10^{15}$ eV
suggests that the longitudinal profile of EAS induced by $\nu_e$'s
is sensibly enhanced with respect to the corresponding proton
showers. On the contrary a large suppression of the profile is
observed for primary $\nu_{\mu}$'s. As a consequence of this, the
threshold for $\nu_e$ shower detection both for the surface array
and the fluorescence detection is lower than for the proton one,
while for $\nu_\mu$ is higher.

This suggests to better explore the possibility to discriminate
between neutrino flavours by the fluorescence detector in the
Auger experiment. In fact, since the depth of the maximum in the
longitudinal development of a shower is directly related to the
primary energy, it should be easy to single out $\nu_e$ induced
showers with CC first interaction at energies lower than the
threshold for detecting NC $\nu_e$ or $\nu_\mu$ showers. Going to
higher energies, an estimate of the NC $\nu_e$ or $\nu_\mu$ shower
contamination can be provided by considering the ratio between
neutrino fluxes of different flavour and the probability of NC
versus CC $\nu$ first interactions.

Future work aims to extend the capabilities of the modified
neutrino version of CORSIKA to energies $\geq 10^{16}$ eV. In this
range one is urged to substitute the simplified hadronic model
HDPM with a more reliable one, like QGSJET. Moreover, the approach
we used for handling unrecognized particle in CORSIKA is no longer
applicable to these high energies, and a treatment of $\tau$
lepton, charmed particles and resonances states, whose interaction
is not included in the present version of the Monte Carlo, is
necessary. Care has to be taken also of the extrapolation of PDF's
to high energy, which determine the value of neutrino
cross-section in the ranges of interest for the Auger experiment.
In fact, while the fraction of events out of the present
acceptable ranges of $x$ and $Q^2$ is 0.02\% at $E_\nu = 10^{15}$
eV and 2\% at $E_\nu = 10^{18}$ eV, it becomes $\sim 40\%$ at
$E_\nu = 10^{21}$ eV. All these research lines are at the moment
under investigation.

\section{Acknowledgements}
We would like to thank D. Heck and J. Knapp for valuable comments
and suggestions. S.P. was supported by the Spanish grant
BFM2002-00345 and a Marie Curie fellowship under contract 
HPMFCT-2002-01831.

\end{document}